\documentstyle[12pt,aasms4,psfig]{article}
\slugcomment{Published by Chin. Phys. Lett. 1998, Vol.15, No.12, p.934}

\begin{document}

\title{The `bare' strange stars might not be bare}
\author{R. X. Xu\altaffilmark{1,2}, G. J. Qiao\altaffilmark{3,1,2}}
\affil{1. CAS-PKU Joint Beijing Astrophysical Center, 
 Beijing 100871, China}
\affil{2. Department of Geophysics, Peking University, 
 Beijing 100871, China}
\affil{3. CCAST (World Laboratory) P.O.Box 8730, Beijing 100080, China}
\altaffiltext{1}{email: rxxu@bac.pku.edu.cn}

\begin{abstract}

It is proposed that the `bare' strange matter stars might not be bare, and 
radio pulsars might be in fact `bare' strange stars. As strange matter stars 
being intensely magnetized rotate, the induced unipolar electric fields 
would be large enough to construct magnetospheres. This situation is very 
similar to that discussed by many authors for rotating neutron stars. Also, 
the strange stars with accretion crusts in binaries could act as X-ray 
pulsars or X-ray bursters. There are some advantages if radio pulsars are 
`bare' strange stars.


\end{abstract}

The first radio pulsar, CP 1919, was discovered in November 1967[1]. Since 
then, more radio pulsars have been found, the number of which is 
approximately 750. However, one of the very interesting and most important 
questions is `What is the nature of pulsars?' which, unfortunately, could 
still not be answered with certainty even now [2].

Soon after the discovery of pulsars, by removing the possibilities of white 
dwarf pulsation and rapid orbital rotation (see, e.g. the review by Smith 
[3]), many people widely accept the concept that pulsars are neutron stars, 
which were conceived as theoretically possible stable structures in 
astrophysics [4,5]. Following this, many authors discussed the inner 
structure of neutron stars, especially the properties of possible quark 
phase in the neutron star core (e.g. Wang \& Lu [6]).

As the hypothesis that strange matter may be the absolute ground state of 
the strong interaction confined state has been raised [7,8], Farhi \& 
Jaffe[9] point out that the energy of strange matter is lower than that 
of matter composed by nucleus for quantum chromodynamical parameters 
within rather wide range. Hence, strange stars should be a possible 
astrophysical object[10], which can be considered as a ground state of 
neutron stars. Pulsars might be strange stars. Therefore, the question 
about the nature of pulsars, which seems to have been answered, rises again.

The important point on this research is to find the difference between the 
behaviors of strange stars and neutron stars, both observationally and 
theoretically. The dynamically damping effects, the minimum rotation 
periods, the cooling curves and the vibratory mode have been discussed in 
detail in the literature. However, no direct observational clue has yet 
shown that pulsars are neutron stars or strange stars.

Almost all of the proposed strange star models for pulsars have addressed 
the case generally contemplated by most authors that the strange star core 
is surrounded by a normal matter crust[11]. The essential features of this 
core-crust structure is that the normal hadron crust with $\sim 10^{-
5}M_\odot$ and the strange quark matter core with mass of $\sim 1.4 M_\odot$ 
and radius of $\sim 10^6$ cm are divided by an electric gap ($>200$ fm).

It is accepted [10] that a strange star with a bare surface will not supply 
charged particles to form a rotating space charged magnetosphere because 
the maximum electric field induced by a rotating magnetized dipole, $\sim 
10^{11}$ V cm$^{-1}$, is negligible when compared with the electric field 
at the strange matter surface, $\sim 10^{17}$ V cm$^{-1}$. Hence bare 
strange stars should not be pulsars. However, by computation, the electric 
field due to electron distribution near the surface decreases quickly 
outward, from $\sim 10^{17}$ V cm$^{-1}$ at the surface to $\sim 10^{10}$ 
V cm$^{-1}$ at a height of $10^{-7}$ cm above the surface (Fig. 1). 
Therefore, a magnetosphere could be established around a bare strange star 
(see below for a detailed discussion).

For a static and nonmagnetized strange star, the properties of strange 
quark matter are determined by the thermodynamic potentials $\Omega_i$ (i 
= u, d, s, e) which are functions of chemical potential $\mu_i$ as well 
as the strange quark mass, $m_s$, and the strong interaction coupling 
constant $\alpha_c$ [10]. For a strange star with a typical pulsar mass 
1.4 M$_\odot$, the total energy $\rho$ has a very modest variation with 
radial distance of strange star [10], from $\sim4\times10^{14}$g cm$^{-3}$ 
(near surface) to $\sim 8\times 10^{14}$g cm$^{-3}$ (near center), 
therefore the quark charge density $\rho_q$ would be order of $10^{15}$ 
($\alpha_c = 0.3$) to $10^{16}$ ($\alpha_c = 0$) Coulomb cm$^{-3}$. 
Physically, as the Fermi energy of quarks becomes higher (for lager $\rho$), 
the effect due to $m_s \not= 0$ would be less important, hence, the charge 
density should be smaller.

For a strongly magnetized and rapidly rotating strange star, the space 
charge separated density near the star is [12]
$$
\rho_{GJ} \sim {1\over 9}\times 10^{-7} B_{12} P^{-1}\;\; {\rm Coulomb}\; 
{\rm cm^{-3}},
$$
where $B_{12} = B_p/(10^{12} {\rm gauss})$ ($B_p$ is the polar cap magnetic 
field), and $P$ in unit of second. As $\rho_{GJ} (\sim 10^{-8})$ is very 
small compared with the quark charge density $\rho_q$ ($\sim 10^{15}$, see 
the Appendix) in the strange star interior, it is a good approximation to 
think the quarks and electrons are in chemical and thermal equilibrium 
although charged particles have been slightly separated to balance the 
unipolar induced electric force in the star interior.

If we have vacuum outside the strange star, the induced electric field along 
the magnetic field, ${\bf E}\cdot{\bf n}_B$, is given by [12]
$$
{\bf E}\cdot{\bf n}_B \sim -1.26\times 10^{11}R_6 B_{12} P^{-1} \cdot 
({R\over r})^4 {\cos^3\theta \over \sqrt{1+3\cos^2\theta}}\;\; {\rm Volt 
\; cm}^{-1},
$$
where a dipole magnetic field is assumed, $B = {1\over 2}B_p 
\sqrt{1+3\cos^2\theta} ({R\over r})^3$, $r$ and $\theta$ are the usual 
polar coordinates with $\theta$ measured from the rotation axis, $R$ is 
the strange star radius, $R_6 = R/(10^6 {\rm cm})$, ${\bf n}_B$ is the 
direction of magnetic field.

In Fig.1, the electric field near $z = 6\times 10^{-8}$ cm to bound the 
\begin{figure}[h]
\centerline{\psfig{figure=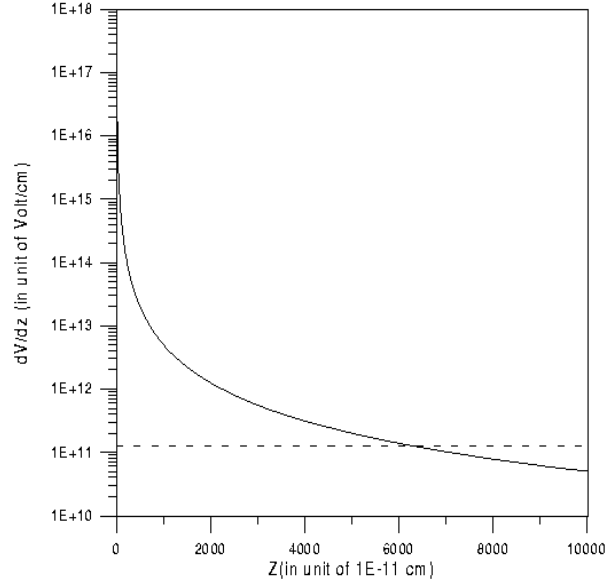,angle=0,height=8cm,width=8cm}}
\caption{
The electric field 
variation curves as function of $z$ (a
space coordinate measuring height above the quark surface).
The dashed lines are for the unipolar induced electric
field.
}
\end{figure}
electrons to the quark matter, dV/dz, is comparable to the unipolar induced 
electric field along ${\bf n}_B$. When $z > 6 \times 10^{-8}$ cm, the motion 
and distribution of electron should be mainly controlled by ${\bf E} \cdot 
{\bf n}_B$, as all of the other forces (e.g. gravitation and centrifugal 
acceleration) can be negligible. Thus, the distributed electrons near and 
above $z > 6 \times 10^{-8}$ cm could not be in mechanically or quantum 
mechanically equilibrium (see the detailed discussion below), and a 
magnetosphere around strange star could be established. 

As discussed above, the unipolar induced electric field does have 
considerable contribution to the distribution of electrons almost at the 
strange star surface ($10^{-8}$ cm is a very small number in astronomical 
viewpoint). The induced field can pull or push the electrons near the 
strange star surface. The potential difference between $\theta = 0^o$ and 
$\theta = 90^o$ is given by[12]
$$
\triangle\phi = 3 \times 10^{16} B_{12}R_6^2 P^{-1} \;\; {\rm volts},
$$
if there are no charged particles outside the star. An electron in this 
electric field can not be continually accelerated, as the pair creation 
processes could stop the acceleration. Electrons or positrons with large 
Lorentz factors should produce $\gamma$ rays by, for example, inverse 
Compton scattering, curvature radiation, synchrotron radiation and, 
perhaps, pair annihilation. Also most of the produced high energy $\gamma$ 
rays in such strong magnetic field could convert to electron-positron pairs 
through $\gamma + B \rightarrow e^{\pm} + B$ or two photon processes, such 
as $\gamma + X \rightarrow e^{\pm}$. Hence, if a strange star have a vacuum 
outside, this cascade of pair creation should bring about the appearance 
of a large enough pair plasma to construct a charge separated magnetosphere 
around the strange star, both in the corotation part and in the open field 
lines [13]. According to energy conservation law, the energy of the above 
cascade process is from the strange star's rotation. As long as the 
magnetosphere has been established, the detailed discussed pulsar models, 
such as the polar gap model [14], the slot gap model [15], and the outer 
gap model [16], would work for the radio as well as higher energy photons' 
emission due to the electromotive force caused by the potential difference 
between the center and the edge of the polar cap region.

Let's come to some details. For ${\bf \Omega}\cdot{\bf B} > 0$, the induced 
electric field would pull the electrons out, accelerates them ultra-
relativistically. As electrons are lost from the quark matter, the strange 
star could be positively charged; the global electric circuit [17] might 
be constructed. However, this global circuit could be in quasi-equilibrium 
and a small vacuum gap similar to that of RS model [14] could be possible 
near the polar cap. For ${\bf \Omega}\cdot{\bf B} < 0$, the induced electric 
field would push the electrons inward, and a large vacuum region should 
be above the polar cap. Some physical processes, such as cosmic 
$\gamma-$rays interaction with strong magnetic fields, or electrons 
(scattered by neutrinos from strange star) synchrotron radiation (jump 
between two Landau levels) could trigger a pair creation cascade.

If a strange star forms soon after a supernova, a magnetosphere composed 
by ions would not be possible since a lot of very energetic outward 
particles and photons are near the star. The time scale T to form an $e^\pm$ 
pair plasma magnetosphere is very small. The total number of $e^\pm$ pairs 
in the magnetosphere might be estimated as $N_{GJ}$,
$$
N_{GJ}\sim R\int_R^{R_c} {\Omega B_p \over 2 \pi c} ({R\over r})^3 rdr 
\approx 7\times 10^{28} R_6^3 B_{12} P^{-1},
$$
where the radius of light cylinder $R_c = {cP\over 2\pi} \gg R$ (the radius 
of strange star). The mean free path $l_p$ of a photon with energy greater 
than $\sim 1$ MeV moving through a region of magnetic B can be estimated 
as [17]
$$
l_p \sim 8.8\times 10^3 e^{3\over 4\chi}/(B \sin\alpha)\;\; {\rm cm},
$$
where $\chi$ could be approximated as 1/15[14], $\alpha$ is the angle 
between the direction of propagation photon and the magnetic field. $l_p 
\sim 68$ cm for $\sin\theta \sim {1\over \gamma} \sim 10^{-5}$ ($\gamma$ 
is the Lorentz factor of electron). Also, the mean free length $l_e$ of 
electron to produce photon by curvature radiation etc. could be assumed 
in order of $l_p$. For the cascade processes discussed, the time scale $T$ 
to build up a magnetosphere might be
$$
T = ({l_p+l_e\over c}){{\rm ln}N_{GJ}\over {\rm ln}2},
$$
which is in order of $10^{-7}$ seconds for typical pulsar parameters. 
Considering the photon escaping and the global magnetic field structure, 
this time scale should not change significantly. 

While strange stars are in binaries, the accretion pressure of wind or 
matter could be greater than the outward pressure, and a strange star could 
be an accretion powered X-ray source. For an accreting strange star, there 
could be two envelope crusts shielding the two polar caps because of 
accretion. As strange matter does not react with ions because of the Coulomb 
barrier (the height of which is $\sim 15$ MeV), there could be an 
electrostatic gap of thickness hundreds fm above the surface.

If the magnetic field is very strong ($B_p \sim 10^{12}$ gauss, High Mass 
X ray Binaries), those accreted crusts should be small. Because of the huge 
release of gravitational energy and the thermal nuclear reactions above 
the crusts, the ion penetration probability might be large enough to keep 
a quasi-equilibrium accretion process, and the accretion-powered strange 
star could be an X-ray pulsar [18].

If the magnetic field is less strong ($B_p \sim 10^8$ gauss, Low Mass X-ray 
Binaries), the accreted polar crusts could be large enough to form a united 
crust. In this case, the accretion process is mild, and the electric gap 
could prevent strong interactions between the crust and the strange matter 
for a long time. However, as the accreted matter piled up the crust could 
be hot enough and dense enough to trigger the thermonuclear flash, this 
strange star could act as an X-ray burster[19], and a lot of ions might 
also be pushed to the strange matter through the Coulomb barrier.

The `bare' strange stars (rather than neutron stars or strange stars with 
crusts) being chosen as the interior of pulsars have some advantages:

1. $^{56}Fe$ emission lines. An iron emission line at 6.4 keV has been 
observed in many accretion X-ray pulsars, while, they have never been 
reported in the seven $\gamma-$ray pulsars. If pulsars' interiors are 
neutron stars, and iron ions could be easily pulled out, there should be 
un-negligible composition of iron in the magnetosphere. Hence, it is hard 
to explain why ion lines have not been observed. However, the open field 
lines region of a strange star magnetosphere consists of e$^\pm$ pairs, 
no ion can result in the radiative processes.

2. Binding energy problem. The RS model [14] is well known to pulsar 
astrophysics both in observation and in theory. Observations show that the 
radio emission might radiate from near the polar cap, to which the RS inner 
gap is related. However, the iron binding energy can not be large enough 
to support the RS gap of a neutron star. Many authors (e. g., Neuhauser 
et al. [20]) used different techniques to estimate the binding energy of 
iron atoms at the neutron star surface. All of them yield the result that 
iron, has its lowest energy state as unbound atoms rather than the chains. 
In the case of a strange star, as the positive charged quark matter near 
the surface is held by strong interaction, the binding energy should be 
approximately infinity when ${\bf \Omega}\cdot {\bf B} < 0$. Because of 
the attraction of the quark matter and the quasi-equilibrium of electric 
current, the electrons could not be easily pulled out when ${\bf 
\Omega}\cdot {\bf B} > 0$.

3. Supernovae explosion. In the collapse of a supernova core, the phase 
transitions from nuclear matter to two-flavor quark matter and from 
two-flavor quark matter to three-quark matter may occur[21], which could 
result in the enhancement of both the probability of success for supernova 
explosion and the energy of the revived shock wave.

A critical point to distinguish neutron star and strange star might be that 
strange stars have very high Coulomb barrier which can support a large body 
of matter, while, neutron stars do not have. As the bare strange stars 
acting as the radio or ( ray pulsars are very similar to the neutron stars 
(the differences of rotation period and cooling curve between them are hard 
to be found), we suggest to search the differences between strange star 
and neutron star in accretion binaries, especially for the bursting X ray 
pulsar GRO J1744-28 [22]. 

\noindent
{\bf Acknowledgements} We are very grateful to Prof. T. Lu and Q. H. Peng for 
their valuable discussion and encouragement. This work is supported by 
National Natural Science Foundation (19673001), by the Climbing project, 
and by the Youth Science Foundation of Peking University.

\end{document}